\documentclass{iau}
\usepackage{graphicx}

\title[IAUS291.~~Temperatures of magnetars] 
{The surface and inner temperatures of magnetars
} 

\author[Z. F. Gao, N. Wang and  Q. H. Peng ]
{Z. F. Gao $^1$, N. Wang $^1$ and  Q. H. Peng $^2$}  
\affiliation{$^1$Xinjiang Astronomical Observatory, CAS, 150, Science 1-Street,\\ Urumqi Xinjiang, 830011, China  \\email: {zhifu$_{-}$gao@xao.ac.cn} \\[\affilskip]
$^2$Shchool of Astronomy and Space Science, Nanjing University, \\Nanjing Jiangshu, 210093, China \\ email: {qhpeng@nju.edu.cn}}

\pubyear{2012}
\volume{291}  
\jname{\mbox{Neutron Stars and Pulsars: Challenges and Opportunities after 80 years}}
\editors{J. van Leeuwen, ed.}
\begin{document}

\maketitle

\begin{abstract}
Assuming that the timescale of the magnetic field decay is approximately
equal to that of the stellar cooling via neutrino emission, we obtain a
one-to-one relationship between the effective surface thermal temperature
and the inner temperature.  The ratio of the effective neutrino luminosity
to the effective X-ray luminosity decreases with decaying  magnetic field.
\end{abstract}


\firstsection 
\section{Introduction}
Magnetars are ultra-magnetized neutron stars (NSs) with magnetic fields
largely in excess of the quantum critical field. The majority
of magnetars are classified into two  populations: the soft
gamma-ray repeaters (SGRs), and the anomalous X-ray pulsars (AXPs). Pulsars
have been recognized to be normal neutron stars, but sometimes have been
argued to be quark stars (e.g., \cite[Xu(2005)]{Xu05};
\cite[Du et al.(2009)]{Du_etal09}). After their formation, magnetars cool
much more efficiently by interior neutrino
emission than by surface photon emission. The neutrino emission mechanisms in the
stellar cores may be divided into two groups, which leads to standard or rapid
cooling. The standard cooling goes mainly via the modified Urca process and the
nucleon-nucleon bremsstrahlung process (e.g.,\cite[Yakovlev \& Pethick(2004)]
{YakovlevPethick04}), whereas rapid cooling is strongly enhanced by the direct
Urca process.

In this paper, we focus on the nonthermal neutrino energy losses in the cores
which control cooling of young and middle age ($t\leq 10^{4}$ yrs, and  $B\sim
10^{14}-10^{15}$ G) magnetars. In our model, for simplicity, we restrict
ourselves by consideration of magnetars whose cores contain the standard
composition of dense matter (neutrons, with some admixtures of protons and
electrons). In the central region of a magnetar, the electron capture process is
expected to occur because of high value of the electron
Fermi energy (e.g., \cite[Gao et al.(2011a)]{Gao_etal11a}; \cite[Gao et al.(2011b)]{Gao_etal11b}).
We calculate the effective neutrino luminosity, and simulate numerically the
relationship between the surface thermal temperature and the inner temperature of
a magnetar.
\section{Surface temperatures of magnetars}
In this section, what we care about is the effective soft X-ray/gamma-ray luminosity
$L_{\rm X}^{\rm eff}$  and the effective surface temperature $T_{\rm suf}^{\rm eff}$
of a magnetar. These two qualities are measured in a local magnetar reference frame.
The effective surface temperature is defined by the Stefan law,
\begin{equation}
L_{\gamma}^{\rm eff}\simeq L_{\rm X}^{\rm eff}=4\pi R^{2}\sigma (T_{\rm suf}^{\rm eff})^{4},
  \end{equation}
where $R$ is the circumferential stellar radius, $\sigma$ is the Stefan-Boltzmann
constant, and $L_{\gamma}^{\rm eff}$ is the thermal surface luminosity
in a local magnetar reference frame.
By using Eq.(1), we plot the diagram of  $Lg T_{\rm suf}^{\rm eff}$ vs. $Lg B$, as shown in Fig.1.
\begin{figure}[t]
\begin{center}
 \includegraphics[width=3.4in]{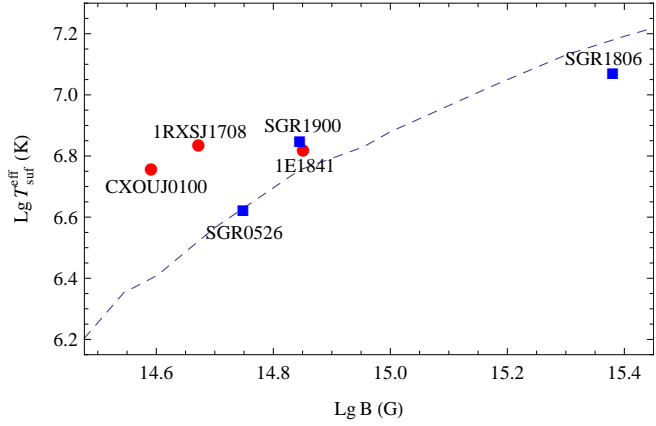}
 \caption{The effective surface thermal temperature $T_{\rm suf}^{\rm eff}$
as a function of $B$. Circles and squares represent AXPs and SGRS,
respectively.}
   \label{fig1}
\end{center}
\end{figure}
\section{Inner temperatures of magnetars}
\begin{figure}[t]
\begin{center}
 \includegraphics[width=3.4in]{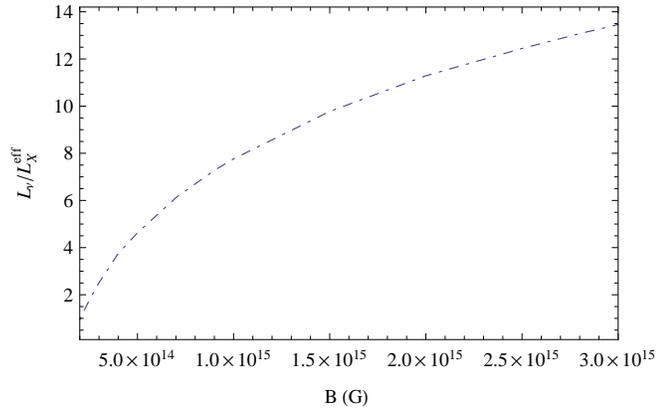}
 \caption{he schematic diagram of $L_{\nu}^{\rm eff}/L_{\rm X}^{\rm eff}$
(or $L_{\nu}^{\infty}/L_{\rm X}^{\infty}$) as a function of $B$.}
   \label{fig2}
\end{center}
\end{figure}

\begin{figure}[t]
\begin{center}
 \includegraphics[width=3.4in]{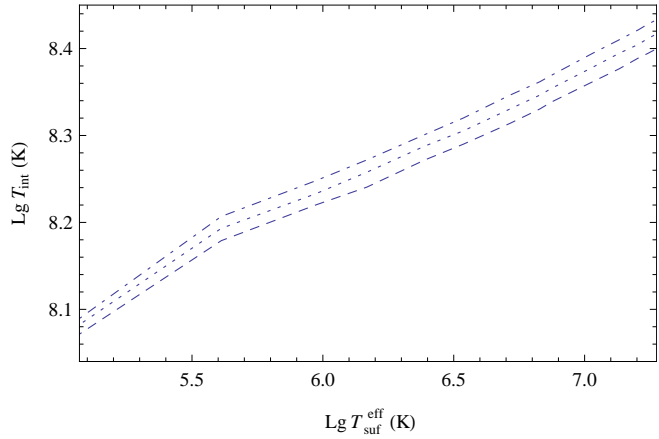}
 \caption{The schematic diagrams of  $T_{\rm int}^{\rm eff}$ vs.
  $T_{\rm suf}^{\rm eff}$.
 The range of $T_{\rm suf}^{\rm eff}$ is assumed to be $(1.835 \times 10^{7}
\sim  1.180 \times 10^{5})$ K arbitrarily, corresponding to $B \sim (3.0\times
10^{15}\sim 1.6 \times 10^{14})$ G. Dot-dashed line, dotted line and dashed line  are
for the initial value of $T_{\rm int}^{\rm eff} = 2.70 \times 10^{8}$ K, $2.60 \times 10^{8}$ K
and $2.50 \times 10^{8}$ K, respectively.}
   \label{fig3}
\end{center}
\end{figure}

According to our magnetar model, we can compute the effective neutrino
luminosity of a magnetars follows:
\begin{eqnarray}
&&L_{\nu}^{\rm eff}=  \Lambda (B,T) V({}^3P_2)\times \frac{(2\pi)^{4}}{\hbar V_1}G_{F}^{2}C_{V}^{2}(1+3a^{2})\nonumber\\
&&\times \int d^{3}n_{\rm e}d^{3}n_{\rm p}d^{3}n_{\rm n}d^{3}n_{\nu}
\delta(E_{\nu}+Q-E_{\rm e}) \delta^{3}(\overrightarrow{K_{f}}- \overrightarrow{K_{i}})S\langle E_{\nu}\rangle,
\end{eqnarray}
where $\Lambda (B, T)$ is the `Landau level-superfluid modified factor',
$S = f_{\rm e}f_{\rm p}(1-f_{\rm n})(1-f_{\nu})$ ($f(\rm j)$
is the fraction of phase space occupied at energy $E_{\rm j}$), and the rest terms are defined in
our recent papers (e.g., \cite[Gao et al.(2011a)]
{Gao_etal11a};\cite[Gao et al.(2011b)]{Gao_etal11b}\cite[Gao et al.(2012a)]
{Gao_etal12a};\cite[Gao et al.(2012b)]{Gao_etal12b}).
We calculate the ratios of $L_{\nu}^{eff}/L_{\rm X}^{eff}$
(or $L_{\nu}^{\infty}/L_{\rm X}^{\infty}$) in different intense fields.
The main results are presented as follows: When the magnetic field $B\sim (3.0\times
10^{15}- 2.0\times 10^{14}$) G, accordingly, $L_{\nu}^{\rm eff}/L_{\rm X}^{\rm eff}$
(or $L_{\nu}^{\infty}/L_{\rm X}^{\infty}$)$\sim 22.93 \sim 1.61$. The details
are to see in Fig. \,\ref{fig2}.
Assuming that the timescale of the magnetic field decay is equal to the
the timescale of stellar cooling via neutrino emission, we obtain a one-to-one
relationship between $T_{\rm suf}^{\rm eff}$ and $T_{\rm int}^{\rm eff}$, shown as in Fig.\,\ref{fig3}.
 \section{Conclusions}
Calculations show that  $T_{\rm int}^{\rm eff}$ is  1-2 orders of magnitude
higher than  $T_{\rm suf}^{\rm eff}$, the ratio of the magnetar neutrino
luminosity to the magnetar soft X-ray luminosity, decreases with decaying magnetic field.

\section{Acknowledgements}
This work is partly supported by Chinese National Science Foundation through grant
No.10773005, China Ministry of Science and Technology under State Key Development
Program for Basic Research (2012CB821800), Knowledge Innovation Program of CAS
KJCX2-YW -T09, the Key Directional Project of CAS and NSFC under
projects 10173020, 10673021, 10773005, 10778631 and 10903019.


\begin{thebibliography}{}
\bibitem[Du et al. (2009)]{Du_etal09}
{Du, Y.J., Xu, R. X., Qiao, G. J.,\& Han, J. L.} 2009, \textit{MNRAS}, 399, 1587
\bibitem[Gao et~al.(2011a)]{Gao_etal11a} {Gao, Z. F., Wang, N., Yuan, J. P., Jiang, L.,\& Song, D. L.}2011, \textit{ApSS},332, 129
\bibitem[Gao et~al.(2011b)]{Gao_etal11b} {Gao, Z. F., Peng, Q. H., Wang, N., Chou, C.-K.,\&  Huo, W. S.}2011, \textit{ApSS},336, 427
\bibitem[Gao et~al.(2012a)]{Gao_etal12a} {Gao, Z. F., Peng, Q. H., Wang, N., \& Chou, C.-K.}2012, \textit{Chinese Physics B}, Vol.21. 5, 57109
\bibitem[Gao et~al.(2012b)]{Gao_etal12b} {Gao, Z. F., Peng, Q. H., Wang, N.,\& Chou, C.-K.}2012, \textit{ApSS}, 342,55
\bibitem[Xu(2005)]{Xu05}{Xu, R. X.} 2005, \textit{MNRAS},356, 359
\bibitem[Yakovlev \& Pethick(2004)]{Yakovlev04} {Yakovlev, D. G., \& Pethick, C. J.}2004, {ARAA} 42, 169
\end{thebibliography}
\end{document}